\documentclass[12pt]{article}
\usepackage[usenames]{color} 
\usepackage{amssymb} 
\usepackage{amsmath} 
  \usepackage{epsfig}

\usepackage{lettrine} 
\usepackage[french]{babel}

\begin{document}

\begin{center}
    {\bf \LARGE   The Impact of Simulations in Cosmology and Galaxy Formation         }\\[4ex]
    {\Large  A summary of the Workshop NOVICOSMO 2008.   }\\[4ex]
 
\parskip=1ex \parindent=3ex     
\vspace {-0.7cm}
 
  {\sl held in SISSA, Trieste, 20-23 October 2008. }

\end{center}
\vspace{0.6cm} 
  
 \abstract{In the study of the process of cosmic structure  formation numerical simulations are     crucial tools to interface observational data to theoretical models and to investigate issues otherwise unexplored.    Enormous advances have been achieved in the last years thanks to the availability of sophisticated  codes. The ever improving performances of large supercomputing facilities coupled with this efficiency of   codes are now allowing to tackle the problem of cosmic structure formation  and subsequent evolution by covering larger and larger dynamical ranges and to  provide  a progressively more realistic account of related complex astrophysical and cosmological processes. Moreover, computational cosmology is the ideal interpretative framework for the overwhelming amount of new data from   extragalactic surveys and from  large sample of  individual objects. The Workshop Novicosmo 2008 "The Impact of Simulations in Cosmology and Galaxy Formation' held  in SISSA was  aimed at providing the state-of-the-art on the latest numerical simulations in Cosmology and in Galaxy Formation. Particular emphasis  was given to the implementation of new physical processes in simulation codes,  to the comparison between different codes and numerical schemes and  how to use  best  supercomputing facilities of the next generation. Finally, the impact on our knowledge on  the Physics of the Universe brought by this new channel of investigation has also  been focused. The Workshop was divided  in  three sections corresponding (roughly) to  three main areas of study: Reionization and Intergalactic medium;  Dark and Luminous matter in galaxies;  Clusters of galaxies and Large scale Structures.  This paper   will   provide  i) a short resume' of the  scientific results of the  Workshop ii) the complete  list of the talks and  the instructions on how  to  retrieve the .pdf of the related   (powerpoint)  presentations iii) a  brief presentation  of the associated Exhibition  "Space Art" 
}

\vspace{0.2cm} 

\section { Highlights of the conference}

It is evident that in different areas  of Cosmology  (N-Body, Hydrodynamical, SPH)  simulations are a new  privileged channel  
to acquire decisive knowledge.

 \subsection {Reionization }
  
With the advent in the near future of radio telescopes as LOFAR, a new window on the high- redshift universe will be opened. In particular, it will be possible, for the first time, to observe the 21cm signal from the diffuse Intergalactic Medium (IGM) prior to its reionization and thus probe the "dark ages".   Simulations of IGM reionization  are already considering its observability.

 \subsection{ Lynam $\alpha$ forest } 
 
The Lyman-alpha forest is  a tracer of intergalactic structures in the high redshift universe.  Hydrodynamical simulations are used to interpret the high and low resolution quasar data sets available and the main physical ingredients that have been incorporated in the numerical codes to properly simulate the transmitted Lyman-alpha flux.  The results  of such simulations  are related to  cosmological parameters and synergy's with other large scale structure observables: they  can measure the  coldness of cold dark matter particles,   non-gaussianities at high redshift and  the intergalactic medium thermal state.

 \subsection {First Stars in the Universe}

The formation of the first stars is thought to occur in very low mass sub-galactic units within the first Gyr of the cosmic history.
The first stars are thought to be extremely luminous and reside in dark matter halos with masses of approximately a million solar masses. These metal-free (or virtually so) stars might have masses of more than 100$M_\odot$, and therefore are expected to end their life as pair-instability/core collapse supernovae. The energy and heavy element deposition by supernovae will affect the star formation rate within the first galaxies and the initial mass function of their stars, via a series of physical processes collectively known as "feedback". Numerical  and radiation hydrodynamics simulations implement the relevant physics to follow in detail these early phases of cosmic evolution and their impact on high-redshift galaxy formation the ensuing chemical enrichment of the IGM  and reionization. The  HII regions created by the first stars are a few kiloparsecs in radius, which then overlap with each other and constitute a volume filling fraction of about a quarter at redshift 15.

\subsection{Clusters}

Clusters of galaxies have been proved to be ideal cosmological probes.
They are
the largest collapsed objects in the Universe and so are very sensitive
to the structure formation process.  Their cosmological simulations  harbor enormous potential for the interpretation of observational data, though they  are extremely challenging, as the structures in and around them span a very large dynamic range in scales. Furthermore, the complexity of the intra cluster medium revealed by multi-frequency observations demonstrates that a variety of physical processes are in action and must be included to produce accurate and realistic models.  In detail,  cosmological hydrodynamical simulations able also  to follow turbulent flows and  subgrid physics,  
study the thermal structure of the intra-cluster medium and its history of chemical enrichment. A particular focus is given   on the effect that feedback from supernovae and AGN has on the observational properties of the hot diffuse cluster baryons,  namely how does it affect   the profiles of entropy, temperature and metallicity and the  scaling relations between X-ray observables.

 \subsection{ Dark Matter Halos} 

N-body simulations of
galactic dark matter halos remain a very hot topic. Ongoing programme
has  simulated a number of  halos at different resolution, including one with 1.4
billion particles corresponding to a particle mass of less than 2000 $M_\odot$
Careful convergence studies validate  the convergence for the structure of the
main halo and its substructures (which number about 300000 in the  largest
simulation). The  nature of the density profile of the main halo, 
its central cusp, the abundance and distribution of subhalos are been clarifying. We are now  able to accurately  resimulate  cosmological initial conditions, 
track dark matter subhalos in the central regions of parent halos, which is essential for understanding the assembly
histories of dark matter halos and 
for running realistic semianalytic models of galaxy formation.
Well defined predictions for the flux of the  gamma-rays arising from the annihilation of
supersymmetric cold dark matter in high density halos have been put forward.

\subsection{Galaxy Structure and  Formation}

Galaxy formation is a  laboratory for testing our knowledge of astrophysics  as good as  cosmology. 
Different kinds of  simulations  are set to reproduce the main  properties   of low- \& high-redshift galaxies and Damped Lyman $\alpha$.  That is   the luminosity/mass   functions, colors and metallicities, stellar content and  molecular hydrogen in the process of establishing the   validity  of  $\Lambda$ CDM scenario.   The physics of how gas accretes into dark matter potential wells and gets converted into stars,  is as crucial as complex. Feedback from stars and AGN and  galactic outflows all  regulate the  complex multi-phase interstellar medium,  in which  magnetic fields, cosmic rays, molecules/dust  play also  a role.
 To reproduce all that  is a  huge computational challenge that requires  to understand  the physics at sub-grid level and also  some  detailed  aspect  of the occurring physical  processes, e.g.   the relation between  the  star formation and the distribution of  molecular hydrogen. Simulations require a post-processing analysis in which  great attention is given  to checks  for numerical artifacts and to pinpoint relevant physical processes. Important phenomenological constraints on the models can be obtained from matching the observed distributions of stellar masses and kinematics Several long standing  open  issues such as the existence of   bulgeless galaxies,  the formation of disks with large angular momentum of the "missing satellites problem" has been tackled  with  a combination of increased numerical resolution and better models for star formation and the energy   balance of the interstellar medium and  detailed star formation histories

\subsection {Gas around Galaxies}

High-resolution  numerical simulations seconded by  a large suite of cosmological gas-dynamical simulations that get repeated using different physical parameters  probed the warm-hot gas halo that is predicted to surround normal disk galaxies in particular the  gas accretion and outflows using  simulations.   The simulations  predict a variety of observational signatures,  including X-ray emission and UV absorption/emission lines, and demonstrate  that Lya, OVI, and CIV spectral lines  are important diagnostics of the strength of feedback in spirals.

 \subsection {AGNs-Coevolution of Black Holes and Galaxies} 
 
The interaction between active galactic  nuclei (AGN) and the intracluster medium (ICM) is crucial in the formation of spheroids. Feedback by hot, underdense bubbles powered by AGN/QSO play a key role in   structure formation.  Hydrodynamical simulations investigate different physical processes claimed to be  responsible for both BH accretion and bulge formation help discriminating among different theoretical models for the quasar lightcurve   and for the
dependence of the quasar lifetime on BH mass.  These simulations  are far from  trivial requiring  modeling of subgrid turbulence and a good understanding of Rayleigh-Taylor   instabilities  to pinpoint  important implications   such as the impact of   AGN-driven clouds on the mixing of metals into the ICM.

\section{Invited and contributed papers}

Here  follows the list of the invited and contributed papers of the Workshop Novicosmo 2008. All the  .pdf files of the related  (power-point) presentations given at the workshop   can be downloaded at  www.novicosmo.org, by clicking Proceedings.
\vskip 0.2cm

\noindent  
\footnotesize {Simulations of Galaxy Formation and Cosmological Reionization  {\it  Renyue Cen} }
 
\noindent 
Feedback processes at Cosmic Dawn: Numerical Views  {\it Andrea Ferrara }

 \noindent  
Starting reionization with the first stars {\it  John Wise }

  \noindent   
Simulations of reionization {\it  Benedetta Ciardi }
 
 \noindent 
Simulating the formation of galaxies and the evolution of the intergalactic medium {\it  Joop Shaye }

 \noindent 
The interaction between galaxies and the intergalactic medium {\it  Tom Theuns }

 \noindent 
The chemical history of the universe {\it  Luca Tornatore }

 \noindent  
The high redshift intergalactic medium as a cosmological probe {\it  Matteo Viel}

 \noindent 
 Simulating the Circumgalactic Medium {\it  Greg Bryan}

\noindent 
Modeling molecular gas and star formation in high resolution Cosmol. simulations {\it A. Kravtsov}
  
 \noindent 
Formation and Evolution of Giant Molecular Clouds in Disk Galaxies {\it Elizabeth Tasker}
 
\noindent 
Galaxy formation simulations in a CDM Universe {\it  Lucio Mayer}

 \noindent 
MUPPI  A new star formation and feedback algorithm for numerical simulations   {\it G. Murante }

 \noindent  
DLAs in Galaxy Formation Simulations  {\it Andrew Pontzen }

 \noindent  
Chemodynamical models of dwarf spheroidal galaxies  {\it Pascale Jablonka }
 
 \noindent 
Simulations of Galaxy Formation Including Outflows  {\it  Romeel Dave'}

 \noindent 
 Modeling Star Formation in Dark Matter Halos  {\it Oleg Gnedin}
 
 \noindent 
 Galaxy kinematics: comparing observations and simulations  {\it Gianfranco Gentile}

  \noindent 
Galaxy Formation Simulations: successes and failures {\it  Kentaro Nagamine}

\noindent 
 Simulations of galactic disks including a dark baryonic component  {\it Yves Revaz}

 \noindent 
The effect of the TPAGB on the semianalytic modeling of galaxy formation  {\it Chiara Tonini}

 \noindent 
Dynamical Downsizing in Ellipticals: clues from Cosmol. Simulations  {\it R. Dominguez-Tenreiro}
 
\noindent 
 Simulation of the AGN ICM interaction  {\it Marcus Brueggen }

\noindent 
Feedback effects on the chemo and thermodynamics of the ICM  {\it Stefano Borgani }

\noindent 
 Magnetic fields, CRs and nonthermal emission in galaxy clusters  {\it  Klaus Dolag }
 
 \noindent 
 Modelling of turbulent flows applied to numerical simulations of galaxy clusters  {\it Luigi Iapichino }
  
  \noindent 
 Coevolution of BHs and Galaxies: models for BH accretion and quasar light curve  {\it  Silvia Bonoli}

\noindent 
  The ICM: Predictions from SemiAnalytic Models of Galaxy Formation  {\it Chris Short }
  
\noindent 
Chemical enrichment in semianalytic models  {\it  Daniel Thomas }

\noindent 
Temperature and mass estimates from simulated Xray galaxy clusters  {\it Riccardo Valdarnini }

\noindent 
Testing scaling relation in situation of extreme mergers   {\it Elena Rasia }

\noindent 
  Second order accurate Lagrangian perturbation initial conditions for cosmological 
resimulations  {\it Adrian Jenkins }
 
 \noindent 
 A New N-Body Simulation of Cosmological Structure Formation {\it  Mike Boylan-Kolchin }

 \noindent 
 The Aquarius project  {\it  Carlos Frenk  }

\begin{figure}
\centering\epsfxsize=9.cm \epsfbox{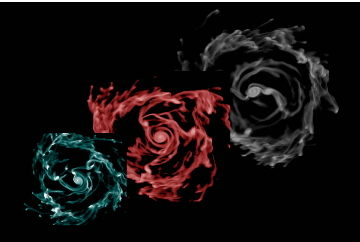}
\vspace{-0.cm} 
\caption{ Image of the movie {\it Virtual Universe} shown at the Exhibition}
 \vspace{-0.3cm} 
\end{figure}

\section {The Exhibition {\it Space Art}.}

Simulations are  "cosmology experiments" leading to important  papers and scientific discoveries but, at the same time, they are    beautiful images and thrilling movies. Once treated by an artist mind,  they   become unique  moments  of  contemplation of  the Cosmos, and of    its mysteries and  a reassessment of the role  of each individual in the Grand Picture of  Nature.   As result, simulations (as astro-images from Large Telescopes) are a  perfect mean to  fill the gap between the curiosity of people and the difficulty of a scientific explanation. Here comes the  Exhibition "Space Art", associated to the Workshop Novicosmo 2008,  in which a number of simulations (many of which performed by participants of  the workshop) are arranged by artists and experts of public outreach   and  presented to the general public.
 The Exhibition, that can be visited at Immaginario Scientifico until 11 January 2009,  includes 4 different paths.  That is   4 multivisions,  displayed on  7 giant screens (3.5m x 2m),  of length of  26', 13', 13' and 20' respectively.   The trailer of the Exhibition is  at  www.youtube.com/watch?v=K0iKLHGqpTo, its  brochure can be retrieved at www.novicosmo.org. Any information, including welcomed enquires for  showing  it around during the Astronomy Year 2009,   should be asked to spaceart08@libero.it.

 {\scriptsize Among the authors of the simulations:  Borgani, Burkert, Coldberg, Dave', Dolag, Dubinsky, Frenk,  Giacomazzo, Gnedin, Gritschneder, Heitsch, Kravtsov, Mayer, Moore, Naab, Nagamine,
    Rezzolla, Schaye, Springer, Theuns, Viel, Wise}
    \noindent
  \vskip .8cm  
   \noindent
\small{ {\bf Paolo Salucci, Stefano Borgani, Carlos Frenk, Lauro Moscardini, Matteo Viel}}

\end{document}